\documentclass[journal=jacsat,manuscript=article]{achemso}

\usepackage[version=3]{mhchem} 
\usepackage{multirow}
\usepackage{xcolor}

\author{Hsin-Mei Ho}
\author{Michael Lorke}
\author{Peter Kratzer}

\email{Peter.Kratzer@uni-due.de}

\affiliation[University of Duisburg-Essen]
{Faculty of Physics, University of Duisburg-Essen, 47057 Duisburg, Germany}

\title[An \textsf{achemso} demo]
  {Quasiparticle level alignment in anthracene--MoS\textsubscript{2} heterostructures}

\begin{document}

\begin{abstract}

Heterostructures composed of transition metal dichalcogenides (TMDCs) and organic molecules have been extensively explored for optoelectronic devices.
To maximize their application potential, it is essential to investigate the electronic band structures, which govern the charge response of the interfaces to external perturbations. 
Based on $GW$ calculations, we present a study of organic-inorganic heterostructures with anthracene molecules adsorbed on monolayer MoS\textsubscript{2}.
Building on previous investigations of organic molecule self-assembly at surfaces, we systematically analyze anthracene configurations with various molecular orientations and surface coverages.
Partially self-consistent $GW_0$ provides qualitatively different level alignments from those in DFT.
Whereas the systems with sparse, horizontally adsorbed anthracenes exhibit type-I alignment, densely packed anthracenes in the head-on position lead to type-II alignment, which indicates the strong dependence of quasiparticle corrections on the interfacial configuration.
These findings highlight the importance of level-alignment predictions for both interpreting experiments and guiding the design of organic-inorganic heterostructures.

\end{abstract}

\section{Introduction}

Stacking different materials through van der Waals (vdW) forces has led to a new paradigm for next-generation optoelectronic devices.
While preserving the advantages of individual components, heterostructures are predicted to exhibit novel properties at the interfaces. \cite{Geim2013}
Transition metal dichalcogenides (TMDCs) have emerged as promising building blocks due to their atomic thickness and interesting light–matter interactions.
On the other hand, oligoacenes such as anthracene and pentacene show their potential in field-effect transistors, organic light-emitting diodes, and solar cells. \cite{Anthony2006}
As candidates for combination with TMDCs, these organic materials offer advantages of chemical versatility and structural flexibility. \cite{Obaidulla2024}

In a heterostructure, how the states around the bandgaps align can profoundly influence the behavior in the presence of external perturbations.
For example, type-II organic-TMDC interfaces are known for interlayer excitons. \cite{Mehdipour2024}
The long-lived charge-separated states have been observed during the relaxation of photoexcited systems via transient absorption spectroscopy. \cite{typeII_Bonkano,typeII_Zhu,typeII_BettisHoman2016}
Therefore, to assess the application potential of multifunctional hybrid systems, it is essential to understand the electronic structures, which leads to the urgent need for a methodology beyond density functional theory (DFT).
While DFT is a ground-state formalism \cite{DFT_Jones1989} and is known for underestimating bandgaps, many-body perturbation theory within the $GW$ approximation offers a more rigorous treatment of quasiparticle energies by describing the screening of the electron–electron interactions. \cite{GW_Hybertsen1986,GW_Reining2017,GW_Krumland2023,Drppel2017}
Notably, the $GW$ approach is not an uniform gap-opening scissor-operator.
Due to differences in orbital characters, quasiparticle corrections in $GW$ approximation can increase the bandgaps to different extents.
In organic-TMDC systems, the electronic properties are primarily determined by the Mo $d$ and C $p$ orbitals, \cite{MoS2_Kadantsev2012,Hummer2005} and $GW$ approximation has been shown to alter the level alignment, compared to plain DFT calculations. \cite{GW_Draxl} 
Since orbital hybridization and electronic screening depend sensitively on the structural arrangement, level alignments of organic-inorganic interfaces may vary according to different self-assembled configurations of molecules, which are often determined by experimental conditions. \cite{structure_Tumino2022,MeyerzuHeringdorf2008,Kim2015}
How the structure of the organic layer changes the electronic properties of an organic-inorganic heterostructure at the $GW$ level remains incompletely understood.

In this work, we aim to fill this gap by examining various anthracene/MoS\textsubscript{2} heterostructures.
We compare the stability of the structures at the DFT level.
Using the $GW_0$ approximation, we investigate the change of the level alignment when anthracene adopts horizontal, vertical orientations and different coverage on MoS\textsubscript{2}.
The importance of performing partially self-consistent $GW_0$ is also discussed.

\section{Results and discussion}

\paragraph{Structures and computational details}

To understand the role of interfacial structures, we define four configurations: F\textsubscript{1}, H\textsubscript{4}, F'\textsubscript{4.5}, and H\textsubscript{18}.

\begin{figure*}
  \includegraphics[width=\textwidth]{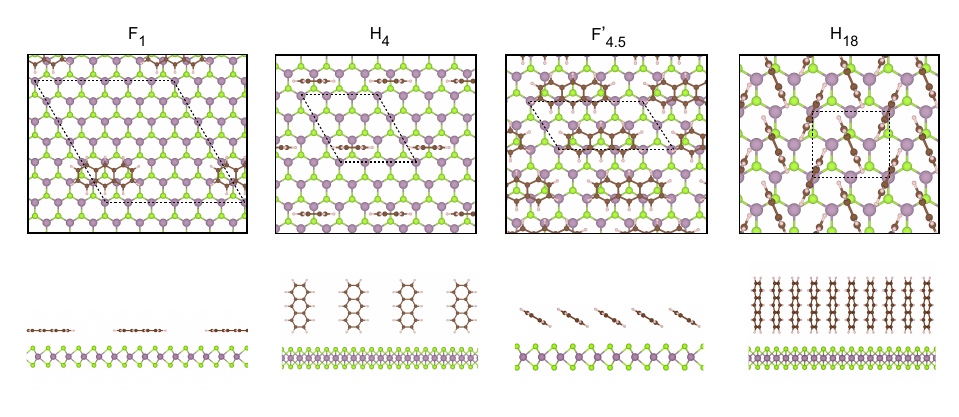}
  \caption{Top views and side views of anthracene/MoS\textsubscript{2} heterostructures with F\textsubscript{1}, H\textsubscript{4}, F'\textsubscript{4.5}, and H\textsubscript{18} configurations.}
  \label{fig:structure}
\end{figure*}

F and H denote the contacts of anthracene molecules with MoS\textsubscript{2}, face-on and head-on.
The indices indicate the equivalent number of anthracenes per $6\times6$ supercell of MoS\textsubscript{2}.  
The sizes of MoS\textsubscript{2} supercells are chosen in order to modify the coverage and consequently the interaction between molecules. 
Illustrated in Figure \ref{fig:structure}, F\textsubscript{1} and H\textsubscript{4} are modeled by hexagonal $6\times6$ and $3\times3$ supercells of MoS\textsubscript{2} monolayer. 
Instead of completely horizontal molecules, the F'\textsubscript{4.5} case is constructed by a $4\times2$ MoS\textsubscript{2} supercell and tilted molecules with an angle to the MoS\textsubscript{2} surface, which is 31{\textdegree} after the structural relaxation.
As for the H\textsubscript{18} case, an orthorhombic $2\times\sqrt{3}$ cell is used to accommodate two vertically oriented anthracene molecules. 
Such herringbone pattern of anthracene crystals has been determined by X-ray diffraction in previous experimental studies. \cite{herringbone}
In each case, a vacuum layer of 10 {\AA} along the out-of-plane direction is included.

We perform DFT calculations using the Vienna ab initio simulation package (VASP.6) \cite{VASP1, VASP2} with plane-wave basis and projector-augmented wave pseudopotential (PAW).
Under the conjugate gradient (CG) algorithm, the atomic structures of all cases are fully relaxed until the interatomic forces are less than $0.01$ eV/{\AA} with vdW interactions DFT-D3 of Grimme taken into account. \cite{vdW-D3}
The lattice constant of MoS\textsubscript{2}, which is calculated and agrees with experiments, \cite{MoS2_Wildervanck1964} is kept at 3.18 {\AA} in all configurations. 
The ground-state properties are calculated with the Generalized Gradient Approximation (GGA) of the Perdew-Burke-Ernzerhof (PBE) \cite{PBE} exchange-correlation functional.
A 400 eV cut-off is applied for the plane-wave expansions of the wavefunctions.
The Brillouin zone is sampled at $\Gamma$ in the F\textsubscript{1} case. 
For cases H\textsubscript{1}, F'\textsubscript{4.5}, and H\textsubscript{18}, $\Gamma$-centered $2\times2$, $3\times3$, and $3\times4$ k-grids are used.
With these choices of the k-grids, we are able to sample the K point of the Brillouin zone of primitive MoS\textsubscript{2}, where the direct bandgap of MoS\textsubscript{2} is located. 
The mapping of k-points is shown in Figure \ref{fig:k-map} of the Supporting information.

Partially self-consistent $GW_0$ calculations are performed in the standard manner: the exchange-correlation functional is first replaced by the many-body self-energy $\mathit{\Sigma=GW}$, which is calculated with the Kohn-Sham energies in the ground-state calculation. 
By solving the quasiparticle equation, we obtain the quasiparticle corrections.
The Green's function $G$ and the electronic self-energy $\mathit{\Sigma}$ are updated self-consistently at each iteration, whereas the screened Coulomb interaction $W$ is kept fixed at its initial value obtained from density functional theory, i.e., $W = W_0$. 
In this scheme, the screening is not iterated, and no vertex corrections beyond the $GW$ approximation are included in either the polarizability or the self-energy.
In the absence of an explicit treatment of vertex corrections---whose consistent implementation remains computationally demanding---this partially self-consistent approach represents the current state of the art for a wide range of semiconductors. 
Empirically, it has been demonstrated to yield quasiparticle energies and band gaps in closest agreement with experiment, outperforming both the single-shot $G_0W_0$ method and fully self-consistent $GW$ calculations that neglect vertex contributions. \cite{GW_Shishkin2006,GW_Shishkin2007_1,GW_Shishkin2007_2,GW_PhysRevB.98.155143}. It should also be noted that, though originally not intended for these, $GW$-type calculations also produce reasonable results for molecules\cite{vanSetten2015}.
The band structure plots are produced via Wannier interpolation by wannier90, \cite{wannier90} and we show the convergence test for $GW_0$ regarding the number of iterations in the Supporting information.

\begin{table}[t]
\caption{Interlayer distances and adsorption energies per molecule}
\begin{tabular}{lcc}
 & Interlayer distance ({\AA}) & $E_\mathrm{ad}$ (eV) \\ \hline
 F\textsubscript{1} & 3.16 & $-1.035$ \\
 H\textsubscript{4} & 2.99 & $-0.232$ \\
 F'\textsubscript{4.5}  & 2.93 & $-0.976$ \\
 H\textsubscript{18}  & 2.77 & $-0.428$
\end{tabular}
\label{table:structure}
\end{table}

We start by examining the relative stability of different configurations before diving into the analysis of the electronic properties. 
We define the adsorption energies $E_\mathrm{ad}$ as the difference between the total energy of the heterostructure and the total energies when anthracene and MoS\textsubscript{2} are separated:
\begin{equation}
  E_\mathrm{ad}=\frac{E_\mathrm{Ant/MoS_2}-N_\mathrm{MoS_2}E_\mathrm{MoS_2}-N_\mathrm{Ant}E_\mathrm{Ant}}{N_\mathrm{Ant}}
\end{equation}
It is normalized by the number of molecules $N_\mathrm{Ant}$ per unit cell. 
$E_\mathrm{MoS_2}$ and $E_\mathrm{Ant}$ are the total energies of primitive MoS\textsubscript{2} and anthracene molecule placed in vacuum.
The negative values in Table \ref{table:structure} suggest that they are energetically stable structures.
This result agrees with previous works that long aromatic molecules prefer being in the flat positions when the overlap between the molecular $\pi$-orbitals and the S $p_z$ orbitals is maximized. \cite{structure_Black2023} 
Head-on adsorption is in general not energetically favorable; however, the total energy is slightly lower when molecules are densely packed in H\textsubscript{18} than in H\textsubscript{4}, which means that the head-on configuration is stabilized via the overlapping $\pi$-orbitals of anthracenes.

\paragraph{Level alignments}

\begin{figure*}[b!]
  \includegraphics[width=\textwidth]{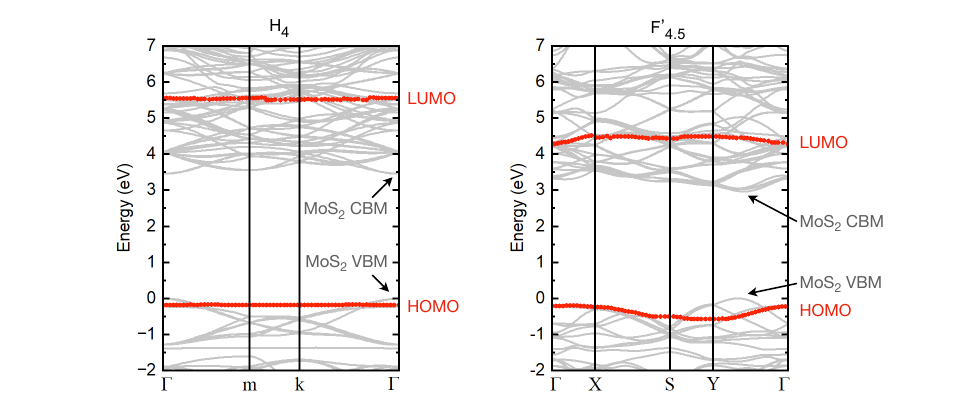}
  \caption{$GW_0$ band structures of anthracene/MoS\textsubscript{2} with H\textsubscript{4} and F'\textsubscript{4.5} configurations. The HOMO and LUMO levels of anthracene are highlighted in red. The Fermi energies are set as 0 eV. The high-symmetry paths $\mathrm{\Gamma-m-k-\Gamma}$ and $\mathrm{\Gamma-X-S-Y-\Gamma}$ are picked from the folded Brillouin zones (Figure \ref{fig:k-map} in the Supporting information). }
  \label{fig:band_compare}
\end{figure*}

As for the electronic properties, we start with the comparison between face-on and head-on anthracenes. 
The $GW_0$ band structures presented in Figure \ref{fig:band_compare} are plotted along the paths $\mathrm{\Gamma-m-k-\Gamma}$ for H\textsubscript{4} and $\mathrm{\Gamma-X-S-Y-\Gamma}$ for F'\textsubscript{4.5}. 
Due to their folded Brillouin zones (Figure \ref{fig:k-map} of the Supporting information), the direct gaps of MoS\textsubscript{2} are located at $\Gamma$ and between $\mathrm{Y}$ and $\Gamma$, respectively.
To obtain the level alignments, we use the orbital characters to identify the positions of the highest occupied molecular orbital (HOMO) and the lowest unoccupied molecular orbital (LUMO) of anthracene.
In both hybrid systems, one can directly find that the LUMO of anthracene is in the conduction band of MoS\textsubscript{2}, while the HOMO is below the MoS\textsubscript{2} valence band maximum (VBM), which suggests the type-I alignment.
Considering the sizes of the MoS\textsubscript{2} supercells, cases H\textsubscript{4} and F'\textsubscript{4.5} have close coverages of anthracene on MoS\textsubscript{2}. 
Namely, at such coverage, the orientation of anthracene does not change the alignment qualitatively.

\begin{figure}[b!]
  \includegraphics[width=8.1cm]{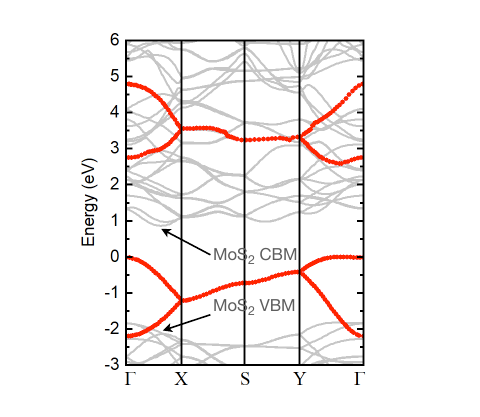}
  \caption{$GW_0$ band structure calculated along the $\Gamma$-X-S-Y-$\Gamma$ path of the orthorhombic cell in the H\textsubscript{18} configuration. The HOMO and LUMO bands of anthracene are highlighted in red. The Fermi energy is set as 0 eV.}
  \label{fig:band_edge-on}
\end{figure}

In Figure \ref{fig:band_compare}, one can notice the small dispersions of anthracene HOMO and LUMO in the F'\textsubscript{4.5} case, which is caused by the interaction between molecules.
Such interaction also affects the HOMO-LUMO gap of anthracene, making it smaller in F'\textsubscript{4.5} than in H\textsubscript{4}.
If we turn our attention to the coverage dependence of the band structure, we can find that these effects of molecular interactions are intense at the high-coverage extreme, H\textsubscript{18}, where anthracene becomes a 2D film.
In Figure \ref{fig:band_edge-on}, the highlighted bands with significant dispersion originate from anthracene HOMO and LUMO.
The direct gap of MoS\textsubscript{2} is folded and located between $\Gamma$ and X, explained in Figure \ref{fig:k-map} of the Supporting information.
Because of the larger electronic screening, the gap of the anthracene layer is significantly reduced.
Essentially, the highest occupied band is now the VBM of anthracene; in other words, the band structure exhibits a type-II alignment.

\begin{figure*}[b!]
  \includegraphics[width=\textwidth]{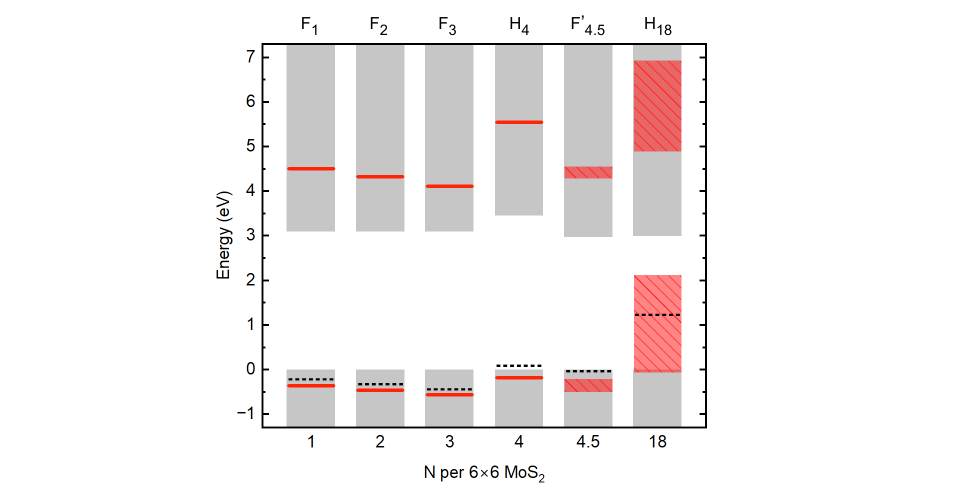}
  \caption{$GW_0$ level alignments between anthracene and MoS\textsubscript{2} with different coverage, N anthracenes per 6$\times$6 supercell of MoS\textsubscript{2}. The VBMs of MoS\textsubscript{2} in all cases are aligned at 0 eV. Cases $N=2$ and 3 (F\textsubscript{2} and F\textsubscript{3}) are created with the interlayer distance of the F\textsubscript{1} case. The HOMOs (tops of HOMO bands for F'\textsubscript{4.5} and H\textsubscript{18}) of anthracene at $G\textsubscript{0}W\textsubscript{0}$ are marked with black dashed lines.}
  \label{fig:coverage}
\end{figure*}

To properly assess the effect of the coverage, we present the level alignments of all cases in Figure \ref{fig:coverage}.
Cases $N=1 \text{, } 4 \text{, } 4.5 \text{, and } 18$ per $6\times6$ supercell of MoS\textsubscript{2} correspond to the already discussed F\textsubscript{1}, H\textsubscript{4}, F'\textsubscript{4.5}, and H\textsubscript{18} geometries.
Shown in Figure \ref{fig:structure_suppl} of the Supporting information, the cases of $N=2 \text{ and } 3$ (F\textsubscript{2} and F\textsubscript{3}) are created by placing horizontal anthracenes on the $6\times6$ MoS\textsubscript{2} supercell with the interlayer distance of F\textsubscript{1}. 
Figure \ref{fig:coverage} clearly shows that at lower coverage, the heterostructures have type-I gap, irrespective of the horizontal or vertical placement of anthracenes.
The energy differences between anthracene HOMO and MoS\textsubscript{2} VBM are within 0.2 to 0.5 eV in these cases.

One should notice that iterating the Green function $G$ in $\mathit{\Sigma=GW}$ can be crucial when the HOMO of anthracene is close to the VBM of MoS\textsubscript{2}.
In Figure \ref{fig:coverage}, the dashed lines indicate the positions of anthracene HOMOs (VBMs in cases F'\textsubscript{4.5} and H\textsubscript{18}) with respect to MoS\textsubscript{2} VBM obtained from $G_0W_0$.
In H\textsubscript{4}, the anthracene HOMO is still above MoS\textsubscript{2} VBM at $G_0W_0$; that is, the level alignment is changed from type-II to type-I when we go beyond $G_0W_0$. 
This suggests that $G_0W_0$ alone could lead to erroneous classification of the interface as being type-II. 
Only a partially self-consistent $GW_0$ gives the qualitatively correct level alignment.

\begin{table*}[t]
\caption{Level alignments of different configurations obtained from DFT-PBE, $G_0W_0$, and $GW_0$: MoS\textsubscript{2} gap, anthracene HOMO-LUMO gap, and the energy difference $\Delta E=E_{\mathrm{Ant.,HOMO}}-E_{\mathrm{MoS_2,VBM}}$. For F'\textsubscript{4.5} and H\textsubscript{18}, $E_{\mathrm{Ant.,HOMO}}$ is taken at the top of the HOMO band.}
\begin{tabular}{lcrrrrrc}

& & F\textsubscript{1} & F\textsubscript{2} & F\textsubscript{3} & H\textsubscript{4} & F'\textsubscript{4.5} & H\textsubscript{18} \\ \hline
 
\multirow{3}{*}{MoS\textsubscript{2} gap (eV)} & PBE & 1.65 & 1.64 & 1.64 & 1.66 & 1.68 & 1.68 \\
& $G_0W_0$ & 2.90 & 2.90 & 2.91 & 3.17 & 2.77 & 2.66 \\
& $GW_0$ & 3.10 & 3.10 & 3.10 & 3.46 & 2.97 & 3.00 \\ \hline

\multirow{3}{*}{Anthracene gap (eV)} & PBE & 2.31 & 2.25 & 2.25 & 2.29 & 2.03 & 1.14 \\
& $G_0W_0$ & 4.48 & 4.42 & 4.46 & 5.21 & 4.10 & 2.63 \\
& $GW_0$ & 4.86 & 4.78 & 4.68 & 5.73 & 4.49 & 2.77 \\ \hline

\multirow{3}{*}{$\Delta E$ (eV)} & PBE & 0.55 & 0.46 & 0.38 & 1.01 & 0.87 & 1.62 \\
& $G_0W_0$ & $-0.22$ & $-0.33$ & $-0.44$ & 0.09 & $-0.03$ & 1.22 \\
& $GW_0$ & $-0.36$ & $-0.46$ & $-0.56$ & $-0.18$ & $-0.21$ & 2.12 \\ 
  
\end{tabular}
\label{table:DFT-band}
\end{table*}

It is also worthwhile to point out that plain DFT calculations fail to describe such coverage-dependent level alignments. 
As we compare the $GW$ results to those by DFT (Table \ref{table:DFT-band}), we find that DFT-PBE only shows type-II alignment regardless of the interfacial configuration.
Consistently, previous research has shown that this limitation is not resolved by the use of hybrid functionals, which yield a type-II alignment for an anthracene/MoS\textsubscript{2} interface with horizontally oriented anthracene. \cite{GW_Krumland2023} 
The quasiparticle corrections not only increase the band gaps of both MoS\textsubscript{2} and anthracene but are also able to capture the changes of environmental screening which are induced by varying the coverage.

\section{Summary and conclusions}

We present a systematic investigation of anthracene-MoS\textsubscript{2} heterostructures with various interfacial geometries. 
We find that DFT gives type-II alignment in all cases and is not able to reflect the changes induced by different orientations and coverages of anthracene.
With partially self-consistent $GW_0$, we conclude that anthracene-MoS\textsubscript{2} generally has a type-I gap.
In the limit of solid anthracene where anthracene molecules are densely packed in the perpendicular orientation, the heterostructure is less energetically favorable. 
The HOMO and LUMO of an isolated anthracene become bands with large dispersion, and type-II alignment is found, in sharp contrast to the face-on cases.
Using the $GW$ approximation, our work underlines the connection between the interfacial structure and the electronic properties of hybrid systems, which is essential for interpreting experiments and guiding the design of organic-TMDCs heterostructures.

\begin{acknowledgement}

This work is funded by the Deutsche Forschungsgemeinschaft (DFG, German Research Foundation) through Project B02 of Collaborative Research Center SFB1242 “Nonequilibrium Dynamics of Condensed Matter in the Time Domain” (Project ID 278162697). 
M.L. acknowledges funding from the DFG under grant number LO 1840/7-1.
The authors gratefully acknowledge the computing time granted by the Center for Computational Sciences and Simulation (CCSS) of the University of Duisburg-Essen and provided on the supercomputer amplitUDE (DFG Grant No. INST 20876/423-1 FUGG) at the Zentrum f{\"u}r Informations- und Mediendienste (ZIM), as well as 
by Zuse Institute Berlin (NHR@ZIB).

\end{acknowledgement}

\newpage
\section{Supporting information}

\renewcommand{\thefigure}{S\arabic{figure}}
\setcounter{figure}{0}

\begin{figure*}[h]
  \includegraphics[width=\textwidth]{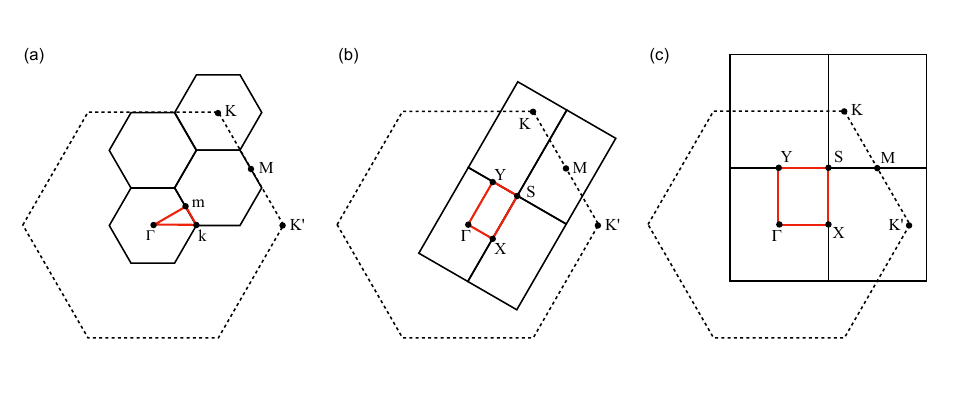}
  \caption{K-point mapping in (a) H\textsubscript{4}, (b) F'\textsubscript{4.5}, and (c) H\textsubscript{18} configurations. The first Brillouin zone of primitive MoS\textsubscript{2} is illustrated by dashed lines, and the folded ones are shown by solid lines. The red lines illustrate the high-symmetry paths along which the band structures are plotted. 
  }
  \label{fig:k-map}
\end{figure*}

\paragraph{Folded Brillouin zones}
The H\textsubscript{4}, F'\textsubscript{4.5}, and H\textsubscript{18} configurations defined in this work consist of $3\times3$, $4\times2$, and $2\times\sqrt{3}$ MoS\textsubscript{2} supercells, respectively. 
Before plotting the band structures, we check how the Brillouin zones are folded. 
In Figure \ref{fig:k-map}, we present the corresponding positions of the K(K') point where the direct gap of a primitive MoS\textsubscript{2} monolayer is located.

\begin{figure*}[h]
  \includegraphics[width=\textwidth]{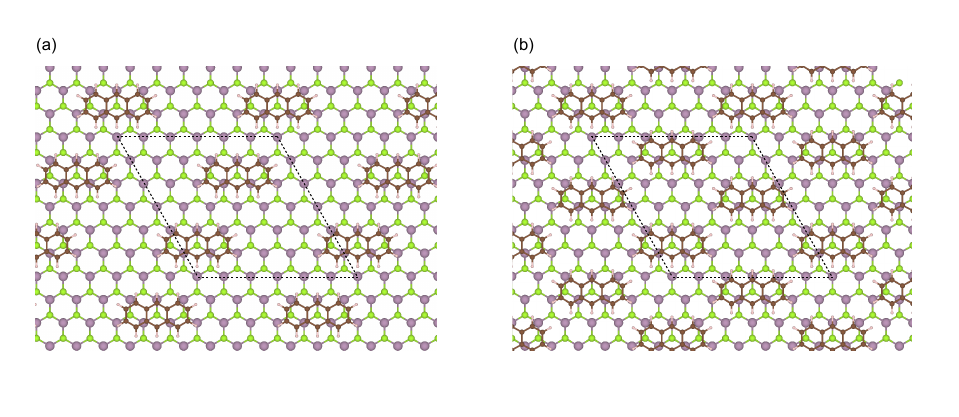}
  \caption{Top views of the heterostructures with (a) F\textsubscript{2} and (b) F\textsubscript{3} configurations, created using the interlayer distance of the F\textsubscript{1} configuration.   
  }
  \label{fig:structure_suppl}
\end{figure*}

\newpage
\paragraph{Structural details}
To complete the investigation of the coverage dependent properties, the cases with two and three horizontally adsorbed anthracenes (F\textsubscript{2} and F\textsubscript{3}) in the $6\times6$ supercell of MoS\textsubscript{2} are constructed, as shown in Figure \ref{fig:structure_suppl}.

\begin{figure*}
  \includegraphics[width=\textwidth]{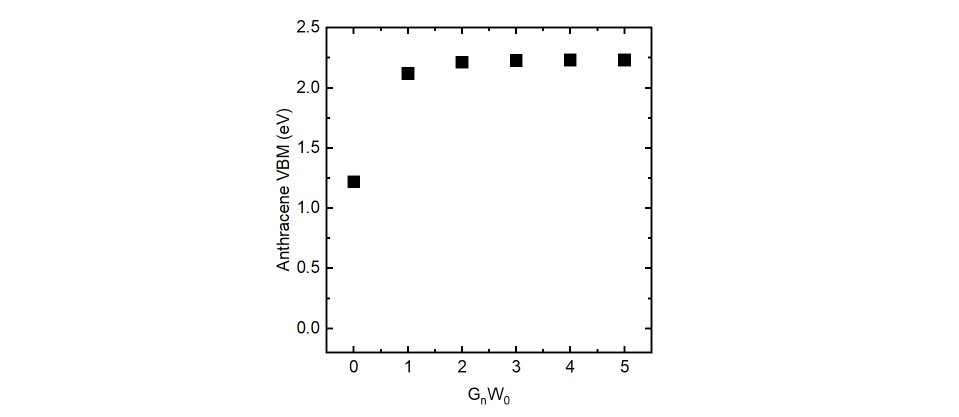}
  \caption{The position of anthracene VBM (top of the HOMO band) calculated by partially self-consistent $GW_0$: from $G_0W_0$ to $G_5W_0$. The MoS\textsubscript{2} VBMs of all cases are moved to 0 eV. 
  }
  \label{fig:G5W0}
\end{figure*}

\newpage
\paragraph{$GW_0$ convergence}
In this work, the $GW_0$ quasiparticle corrections are taken at the $G_1W_0$ level.
We check its validity by iterating the Green function $G$. 
As shown in Figure \ref{fig:G5W0}, $G_1W_0$ is sufficient for the investigation of level alignments. 
Since this convergence test is done for H\textsubscript{18} where the relative position of anthracene VBM has the largest change from $G_0W_0$ to $G_1W_0$ among all heterostructures, we conclude that the calculation of level alignments of such hybrid systems requires at least $G_1W_0$.

\providecommand{\latin}[1]{#1}
\makeatletter
\providecommand{\doi}
  {\begingroup\let\do\@makeother\dospecials
  \catcode`\{=1 \catcode`\}=2 \doi@aux}
\providecommand{\doi@aux}[1]{\endgroup\texttt{#1}}
\makeatother
\providecommand*\mcitethebibliography{\thebibliography}
\csname @ifundefined\endcsname{endmcitethebibliography}  {\let\endmcitethebibliography\endthebibliography}{}

\end{document}